# Einstein's $E = mc^2$ mistakes

## Hans C. Ohanian

*Dept. of Physics, University of Vermont, Burlington, VT 05405-0125, USA*

**Abstract**
Although Einstein's name is closely linked with the celebrated relation $E = mc^2$ between mass and energy, a critical examination of the more than half dozen "proofs" of this relation that Einstein produced over a span of forty years reveals that *all* these proofs suffer from mistakes. Einstein introduced unjustified assumptions, committed fatal errors in logic, or adopted low-speed, restrictive approximations. He never succeeded in producing a valid general proof applicable to a realistic system with arbitrarily large internal speeds. The first such general proof was produced by Max Laue in 1911 (for "closed" systems with a time-independent energy-momentum tensor) and it was generalized by Felix Klein in 1918 (for arbitrary time-dependent "closed" systems).

Keywords: special relativity, mass-energy relation, energy-mass relation, Einstein, Laue, Klein

## 1. Introduction

Einstein's 1905 derivation (Einstein, 1905) of the celebrated relation between mass and energy is clouded by controversy. Not only is there a question of priority—J. J. Thomson, Abraham, Poincaré, and Lorentz had recognized that the electrostatic energy of a charge distribution is endowed with mass, and they had proposed that most or all of the mass of the electron arises from its electric self-energy; and Hasenöhrl had shown that electromagnetic radiation enclosed in a cavity contributes to the inertia of the cavity (Rohrlich, 1965, Chapt. 2; Whittaker, 1960, Vol. I, pp. 309, 310 and Vol II, pp. 51, 52)—but there is also a question of the validity of the argument that Einstein used in his 1905 paper. Planck (1908, p. 29, footnote) objected that Einstein's argument rested on an "assumption permissible only as a first approximation," and Laue (1911) criticized Einstein's use of the nonrelativistic approximation for the internal dynamics of an extended body. Furthermore, in 1952 Ives gave a lengthy analysis of Einstein's derivation from which he concluded that Einstein's argument was logically circular—he claimed that in some step of the argument Einstein had assumed what he was supposed to prove. In the final sentence of his paper, Ives rendered his summary judgment: "The relation $E = mc^2$ was not derived by Einstein" (Ives, 1952). This judgment was strongly supported by Jammer (1961, pp. 177 et seq.) in his well-known book *Concepts of Mass in Classical and Modern Physics* and also by Arzeliès (1966, p. 75 et seq.) and Miller (1981, p. 377) in their later books.

     In 1982 Stachel and Torretti analyzed Ives' analysis, and concluded that Ives was wrong and Einstein was right: "…if we are not willing to countenance some mind-boggling metalogical innovation, we have to declare that Ives, Jammer, and Arzeliès—not Einstein—are guilty of a logical error." They then go on to say that Einstein's "premises are certainly strong enough to derive the mass-energy equivalence relation…" (Stachel and Torretti. 1982).

     I will show that although Stachel-Torretti were right in their criticism of Ives, Jammer, and Arzeliès, they were wrong in accepting Einstein's derivation. The mass-energy formula cannot be derived by Einstein's 1905 argument, except as an approximate relation valid in the limit of low, nonrelativistic velocities for the internal motions of the system under consideration. The defect in Einstein's argument is not a *petitio principii*, but a *non sequitur*. Einstein's mistake lies in an unwarranted extrapolation: he assumed that the rest-mass change he found when using a nonrelativistic, Newtonian approximation for the internal motions of an extended system would be equally valid for relativistic motions. Indeed it is—but Einstein failed to prove that in 1905, and he failed again in all his later attempts. To mend this mistake, Einstein needed to prove that the kinetic energy of an extended system has the exactly the same dependence on velocity as the kinetic energy of a particle. He never proved this, not in the 1905 paper, nor in any subsequent paper. Only in one attempt in 1907, did Einstein produce a valid derivation of the mass-energy

relation, but only for a highly idealized, unrealistic system consisting either of electromagnetic radiation confined in a massless cavity or a massless, electrically charged body.

These early derivations, or attempted derivations, dealt only with special cases, that is, they were really special instances rather than general proofs. The first complete and general proof of $E = mc^2$, valid for an arbitrary closed, static system, was constructed in 1911 by Laue. A more general proof, valid for an arbitrary closed, time-dependent system, was finally formulated in 1918 by the mathematician Felix Klein (Laue, 1911; Klein 1918a).

## 2. The controversy
In his 1905 paper, Einstein examined the change in the translational kinetic energy of an extended body when it emits a pair of light pulses in opposite directions. To determine the implications of this emission process for the rest mass of the body, he needed a definition of the kinetic energy of the body. In the early days of relativity, it was known that the kinetic energy of a particle is

$$K = mc^2 \left( \frac{1}{\sqrt{1 - v^2 / c^2}} - 1 \right),$$ (1)

but it was not thought self-evident that the translational kinetic energy of an extended body  has the same velocity dependence as that of a particle.

In nonrelativistic physics, it is straightforward to prove that the translational kinetic energy of an extended body is like that of a particle moving at the speed of the center of mass. To prove this, we split the sum of the kinetic energies of the particles or mass elements in the body into a translational kinetic energy of the center of mass, and a sum of kinetic energies relative to the center of mass. The former is simply $\frac{1}{2}Mv^2$, where $M$ is the sum of the masses and $v$ the velocity of the center of mass. But this simple split hinges on nonrelativistic kinematics and dynamics (the Galilean addition law for velocity, the choice of the center of mass as fiducial point, and Newton's third law, which ensures that the effects of the internal action and reaction forces cancel so as to give the center of mass a particle-like equation of motion).

In relativistic physics, the derivation of a corresponding result for the translational kinetic energy of an extended body is considerably more difficult. In fact, a naïve analysis of some simple examples suggests that it might not be true [see the example in Section i)], and a general result was not obtained until Laue's work in 1911. To bypass this roadblock, Einstein adopted an indirect definition of the kinetic energy in 1905 that, at least ostensibly, did not seem to require any consideration of the internal dynamics of the extended body. Einstein defined the kinetic energy of an extended body moving with some speed $v$ in some given inertial reference frame as the difference between the energy of the body in that reference frame and the energy of the body in an inertial reference frame in which it is at rest. Stachel-Torretti commended Einstein for this indirect but precise definition and said that Einstein "studiously avoided using it [Eq. (1)] in the derivation of the mass-energy equivalence…for he had as yet no grounds for assuming that the dependence of the kinetic energy on the internal parameters can be summed up in a rest mass term."

With his general definition of kinetic energy, Einstein showed that when a body emits two pulses of light of energy $E/2$ in opposite directions in its rest frame, the change of kinetic energy in some other frame is (in modern notation)

$$K_2 - K_1 = -E \left( \frac{1}{\sqrt{1 - v^2 / c^2}} - 1 \right)$$ (2)

In the next and last step of his argument, Einstein resorts to a low-speed approximation, with $K_1 = \frac{1}{2}m_1 v^2$ and $K_1 = \frac{1}{2}m_2 v^2$. Substituting these approximations into Eq. (2) and comparing terms of order $v^2$, he obtains his mass-energy relation, $m_1 - m_2 = E / c^2$.

Ives, Jammer, and Arzeliès overruled Einstein's definition of the kinetic energy. Blithely accepting the particle formula Eq. (1), they concluded that the change of kinetic energy of the body must be

$$K_2 - K_1 = (m_2 - m_1)c^2 \left( \frac{1}{\sqrt{1 - v^2/c^2}} - 1 \right) \qquad (3)$$

and they claimed that since the right sides of Eqs. (2) and (3) differ by a factor $E/(m_1 - m_2)c^2$, Einstein must have "unwittingly assumed" (Jammer, 1961, p. 179) that $E/(m_1 - m_2)c^2 = 1$, which would indeed imply his argument is logically circular. But, as Stachel-Torretti quite correctly pointed out, Einstein's deduction of Eq. (2) is independent of Eq. (3), and therefore the comparison of these equations does not establish a vicious circle. It merely provides a short cut to the mass-energy equivalence: if *both* Eqs. (2) and (3) are valid, then division of the first of these equations by the second immediately yields $E/(m_1 - m_2)c^2 = 1$ as a mathematical consequence [and there is then no need to go through the next and last step of Einstein's argument, which is designed to avoid the particle formula Eq. (1)].

Thus, Stachel-Torretti were right in asserting that Ives, Jammer, and Arzeliès are wrong. However, Stachel-Torretti were wrong in asserting that Einstein is right. His argument contains three mistakes, not in a vicious circle, but in the last step of his argument and also in the definitions he adopted (or failed to adopt). The three mistakes in Einstein's argument are i) failure to examine the full dependence of Eq. (2) on the velocity $v$; ii) failure to examine the physical basis and the implicit assumptions in the definition of the kinetic energy; iii) failure to provide a definition of the velocity $v$ of the body.

Item i) means that Einstein's derivation is logically incomplete, because, although the lowest-order approximation for the kinetic energy, $K = \frac{1}{2}mv^2$, leads to the Einstein's mass-energy relation, it is not self-evident that this approximation remains valid when the internal motions of the body are relativistic. A proof is required to establish this approximation for the kinetic energy, and that is far from trivial. Items ii) and iii) are not mistakes in logic, physics, or mathematics, but rather deficiencies in exposition (or in propaedeutics). These deficiencies indicate that in 1905 Einstein had misconceptions about the physical rationale for these definitions, and he had an incomplete grasp of the complications inherent in the relativistic dynamics of an extended system. As was often the case in his work, he was navigating through a fog, and he was relying on his superb physical intuition to bring him to a safe port.

i) ***Velocity dependence of the kinetic energy.*** To exhibit the mistake in Einstein's attempt to bypass an explicit examination of the velocity dependence of the kinetic energies in Eq. (2), consider the following simple *Ansatz* for the kinetic energies: $K_1 = m_1 F_1(v)$ and $K_2 = m_2 F_2(v)$, where the functions $F_1$ and $F_2$ are assumed to depend on the internal properties of the body before and after the emission of the light, that is, these functions are *not* universal functions of the velocity.[1] With this *Ansatz*, Eq. (2) becomes

$$m_2 F_2(v) - m_1 F_1(v) = -E \left( \frac{1}{\sqrt{1 - v^2/c^2}} - 1 \right) \qquad (4)$$

If the translational velocity $v$ and also the internal velocities of the particles within the body are low (nonrelativistic), then Newtonian physics is approximately valid, and then the functions $F_1$ and $F_2$ must be approximately the same as for the translational motion of a particle, $F_1 = F_2 = \frac{1}{2}v^2 + \ldots$ Substituting these approximations into Eq. (4) and comparing the terms of order $v^2$, we obtain, of course, Einstein's result $m_1 - m_2 = E/c^2$. We can go a step beyond that if we then substitute this result into Eq. (4), so we obtain

$$m_2 F_2(v) - m_1 F_1(v) = c^2(m_2 - m_1) \left( \frac{1}{\sqrt{1 - v^2/c^2}} - 1 \right) \qquad (5)$$

In this equation we can regard $m_1$ and $m_2$ as independent parameters, and we can therefore conclude that the functions $F_1$ and $F_2$ must necessarily be of the form $F_1 = F_2 = c^2(1/\sqrt{1 - v^2/c^2} - 1)$, that is, we can conclude that these functions are universal, and that they are exactly the same as for a particle. This is an interesting corollary of Einstein's 1905 result, which Einstein somehow overlooked. In essence this says that for a body with low internal velocities, the particle-like behavior of the translational motion persists even when the translational velocity becomes large.[2]

Whether this corollary is judged as supporting or undermining Einstein's proof depends on the context in which it is viewed. If we take for granted the consistency of relativistic mechanics, then this (indirect) derivation of the expression for the kinetic energy of an extended body is a bonus. But in the early days of relativity, the consistency of the relativistic mechanics of extended bodies was in contention, and most physicists would have wished for a direct derivation of the expression for the translational kinetic energy (by introduction of some kind of center of mass velocity, more or less in the manner familiar from Newtonian mechanics), so as to demonstrate explicitly the consistency with the expression extracted from Eq. (5).[3] In the absence of such a demonstration of consistency, Einstein's proof of the mass-energy relation might be judged as incomplete or premature.

Thus, the significance of the corollary for Einstein's proof is somewhat murky. But one thing is clear: Einstein's proof suffers from a fundamental limitation in that it is *not valid* if the internal velocities of the particles within the body are large (relativistic). The trouble is that for large internal velocities, Newtonian physics is not applicable to the internal mechanics of the body, and the Newtonian approximation $\frac{1}{2}mv^2$ for the translational kinetic energy is then not self-evident, even if the body's translational velocity $v$ is low.

Here is a simple example of how and why the kinetic energy of an extended body might not mimic that of a particle and might not be consistent with the Newtonian approximation: Suppose that the extended body consists of some particles confined in a cylindrical massless box within which they bounce back and forth elastically, moving at a constant, high velocity $u$ parallel to the axis of the box, such that at each instant equal numbers of particles are moving forward and backward (a one-dimensional gas). By means of the relativistic addition rules for velocity, it is easy to show that when this system has a translational velocity $v$ in a direction *perpendicular* to its axis, the sum of particle energies in this system is proportional to $1/(1 - v^2/c^2)^{1/2}$, which displays the expected particle-like dependence on the translational velocity.

However, when this system has a translational velocity in a direction *parallel* to its axis, then the sum of particle energies includes not only the expected term proportional to $1/(1 - v^2/c^2)^{1/2}$, but also an unexpected term proportional to $(u^2v^2/c^4)/(1 - v^2/c^2)^{1/2}$, which means that the velocity dependence of the total energy (and thus also the kinetic energy) is *not* particle-like. Taken at face value, the extra term implies that, even at low velocity $v$, the kinetic energy is anisotropic, that is, the kinetic energy for parallel motion deviates from the kinetic energy $\frac{1}{2}mv^2$ for perpendicular motion by an extra amount of approximately $mu^2v^2/c^2$, in contradiction to Newtonian mechanics. [Einstein (1907a, pp. 373-377) later discovered that the mechanical stress in a moving system contributes to the energy. The stress in walls of the box generated by the particle impacts on its ends makes a negative contribution to the total energy when the box has a parallel translational velocity, and this extra contribution exactly cancels the "unexpected" term, so the total energy of the complete, closed system of particles *and* box actually has the expected particle-like form. Another instructive example illustrating this, involving the electric energy in a capacitor, was recently discussed by Medina (2006).]

Einstein's attempt to bypass the construction of an explicit formula for the translational kinetic energy of an extended body was a mistake. This explicit formula is a necessary part of the proof of the mass-energy relation. Without such an explicit formula, the proof is incomplete; at best, it shows that the mass-energy relation is valid in an approximate sense for a body with low, nonrelativistic internal velocities. That is, the energy $E$ is approximately subsumed in the mass $m_1$ of the body when the internal motion is nonrelativistic, but the energy $E$ might be distinguishable from "true" mass if the internal motions are relativistic (or if we measure the mass with high precision so as to reveal even small relativistic effects).

Thus, Einstein's premises are not strong enough for a general derivation of the mass-energy relation. The later work of Laue and Klein has made it clear that Einstein needed two extra ingredients to complete the derivation: the conservation law for the energy-momentum tensor and a precise definition of the velocity $v$ (see below).

ii) *Definition of the kinetic energy.* Although, at first sight, Einstein's indirect definition of the kinetic energy seems incontrovertible, it suffers from two subtle problems, overlooked by Einstein and also by Stachel-Torretti. To calculate the energy, we must integrate the energy density over the volume of the body or system. In the first reference frame, this integration runs over, say, the hypersurface $t = 0$, whereas in the second reference frame, it runs over the hypersurface $t' = 0$. These 3-D hypersufaces slice across the 4-D worldtube of the system at different angles, that is, they capture contributions from different times, when the internal evolution of the system has perhaps brought it into a different state, so it has effectively become a "different" system. Before we blindly accept Einstein's definition, we need to establish that it makes sense to interpret the energy difference between such effectively "different" systems as kinetic energy.

It is immediately obvious that when an extended system is under the action of time-dependent external forces, this interpretation does *not* make sense. For instance, if a force is applied at some remote point of the system, it may happen that in one reference frame this force is acting at time $t = 0$, whereas in the other reference frame it is not yet acting at time $t' = 0$, and it is then nonsensical to pretend that the energies in the two reference frames differ only because of their relative motion, as demanded by Einstein's definition. Thus, when an extended system is under the action of forces, it is *impossible* to define a kinetic energy. In any inertial reference frame, we can define the *total* energy at any instant of time (by integration of the energy density), but we cannot separate out a *kinetic* energy.[4]

A second problem arises if the system is held in equilibrium by opposite external forces, for instance, a volume of gas or of blackbody radiation held in equilibrium by the constant external pressure exerted by the container (which we do not regard as part of the system). Such external forces perform no net work, but they contribute to the energy of the moving system by their stress [as in the example in Section i)]. Planck investigated this problem in detail in an analysis of the total energy of a moving volume of blackbody radiation (Planck 1908), and he concluded that for this system it was not possible to separate the total energy into kinetic energy and rest energy. For a volume of blackbody radiation [and also for a volume of relativistic gas, as in the example in Section i)] that has a low overall translational velocity $v$ but contains individual waves or particles of high velocities, the translational kinetic energy fails to take the expected form $\frac{1}{2}Mv^2$, where $M$ is defined as the ratio of the translational momentum and the velocity, as required by the presumptive validity of Newtonian mechanics at low translational velocity. Therefore the kinetic energy cannot be separated from the total energy, that is, Einstein's definition of the kinetic energy fails because of an inconsistency with the definition of momentum.

To eliminate this inconsistency, it is necessary to take into account the relativistic contributions that the stresses in the walls of the container make to the total kinetic energy and momentum, that is, we need to restrict Einstein's definition of kinetic energy to closed systems that include the sources of all the "external" forces, even if these forces perform no net work. Hence Planck was perfectly correct in his objection that "only in the first [nonrelativistic] approximation" is it possible to assume that "the total energy of a body is an additive combination of its kinetic energy and the energy in the rest frame" (Planck, 1908).[5]

For a system with no external forces (that is, an isolated system, or a "closed" system), it might make sense to interpret the energy difference between the two reference frames as kinetic energy, because energy conservation in the (inertial) rest frame of the system ensures that the work done by internal forces does not contribute to the energy difference. However, to examine this in detail, it is necessary to take into account the relativistic relationships among the energy density, momentum density, and stress. These are all related by the Lorentz transformation equations and by the conservation law for the components of the energy-momentum tensor. Keeping track of all these relationships is not a trivial matter, and to prove that Einstein's subtraction prescription makes sense requires some skill in tensor analysis and 4-D integration (Gauss's theorem in 4 D). In 1905 Einstein was not ready to handle these complications. He did not mention of any of these issues, and at that time he seems to have been completely unaware of the problems associated with his definition of the kinetic energy.

iii) *Definition of the velocity*. The final deficiency in Einstein's argument is the absence of any definition of the velocity $v$ of the system. In 1905 Einstein apparently believed that the meaning of $v$ was self-evident. Maybe he had in mind a nonrotating rigid body, whose velocity can be defined by marking a fiducial point on the body. But if we are dealing with a system with internal motions, we have to be more careful. If the internal motions in the system are nonrelativistic, then the appropriate choice of fiducial point is obvious: it is the center of mass. However, if the internal motions are relativistic or if the system contains a substantial amount of field energy (including potential energy), then a modification of the naïve Newtonian definition of the center of mass is required: instead of the centroid of the (rest) mass distribution, we must use the centroid of the energy distribution.[6] Einstein did not say anything about this until his next paper on the mass-energy relation, in 1906.

### 3. Who proved $E = mc^2$?

Einstein returned to the mass-energy problem in six other papers: one in 1906, two in 1907, an unpublished paper in 1912, and, much later, two more papers in 1935 and 1946 (Einstein, 1906, 1907a, 1907b, 1912, 1935, 1946). These reprises are in themselves an indication that Einstein had some suspicions that his proofs were unsatisfactory—didn't Feynman say that one good proof is sufficient?

In the 1906 paper Einstein analyzes the motion of the center of mass of a system containing several small bodies and electric fields. In that paper, Einstein includes electric field energy in his definition of the center of mass, treating the field energy as a mass distribution, in accord with $E = mc^2$, but he still treats the contribution from the moving bodies as nonrelativistic and considers only what their rest masses contribute to the center-of-mass calculation. This is inconsistent, because if the field energies and the potential energies in a system are large, then so will be the kinetic energies of the charged bodies moving in these fields (in fact, for periodic motions, the virial theorem *demands* that the potential and kinetic energies be of the same order of magnitude). Why does Einstein focus on the contribution of the field energy to the position of the center of mass, but ignores the contribution of the kinetic energies? The answer might be that he tried to include the latter but could not find his way around the obstacles described in ii).

In essence, he is again repeating the mistake of the 1905 paper, by failing to examine the contribution that the detailed velocity dependence of the kinetic energy makes to his calculation. However, he now recognizes that his result is only approximate, saying: "If we ascribe to any energy $E$ the inertial mass $E/V^2$ [ that is, $E/c^2$ in modern notation], then the principle of conservation of the motion of the center mass is valid, at least to first approximation."

In the two 1907 papers Einstein deals with very special kinds of extended systems, subject to various restrictive assumptions. Both of these papers contain a variant of the mistake of the 1905 paper. The first paper deals with an extended system consisting of electric fields and electric charges held in static equilibrium by a rigid mechanical framework. In his analysis of this system, Einstein displays sharp insight into the implications of the mechanical stress that holds the charges in equilibrium, and he performs a pretty calculation to establish that the presence of this stress increases the kinetic energy of the system. Modern tensor techniques make it trivial to establish this result by examination of the Lorentz-transformation properties of the energy-momentum tensor; but, with remarkable virtuosity, Einstein extracts this result from a direct calculation of the work performed during the switch-on of the stress. He finds that the difference between the kinetic energies of systems with charges and without charges is given by an expression of the form of Eq. (2) with a positive sign on the right side, where $E$ now represents the electrostatic energy in the rest-frame of the charged system.

This is an impressive result, but, without knowledge of the mathematical form of the kinetic energy, it does not permit a derivation of the mass-energy equivalence. Einstein now disregards this roadblock—he cavalierly assumes that the kinetic energy of the extended system has the simple particle-like form Eq. (1), and from this he obtains the desired mass-energy relation. This is a surprising about-face from his more cautious approach in earlier years, when he had avoided such an assumption about the kinetic energy of an extended system.

The same is true of the second 1907 paper, which is a modification of the 1905 paper. Instead of changing the energy of the system by emission of light pulses, he changes it by the action of an external electric field arranged in such a way that it removes (or adds) energy, but not momentum. This again gives him, in essence, Eq. (2). He again assumes that the kinetic energy of the system has the simple particle-like form Eq. (1), making the same mistake as in the earlier paper of that year. Thus, all of these attempts to prove the mass-energy equivalence come to grief on his failure to demonstrate that the kinetic energy of an extended system has the same form as that of a particle.

However, there are two exceptional cases, briefly mentioned in the second 1907 paper (Einstein, 1907b, Section 14), for which Einstein's arguments are valid: a system consisting of electromagnetic radiation confined in a massless container and a system consisting of massless electric charges placed on a massless rigid framework (the container or the framework provides the forces needed to hold the system in equilibrium). In these cases, it is possible to pretend that the initial energy is zero, and that the final energy is entirely attributable to the electromagnetic radiation or the electric fields generated by the action of external electric forces on massless, or nearly massless, electric charges in the walls of the container or on the framework. The initial kinetic energy is then zero, and the final kinetic energy is completely determined by the right side of Eq. (2) (again, with an opposite sign), which establishes that the system behaves like a particle of mass $E/c^2$. This derivation was the first complete and valid proof of the mass-energy equivalence, albeit restricted to quite artificial, totally unrealistic systems.

In the 1906 and 1907 papers Einstein displays much virtuosity in the handling of electromagnetic fields, but also a surprising (for him) lack of insight into the deeper aspects of the problem. He never recognized that the real key to the mass-energy relation was the conservation law for the energy-momentum tensor and that all the electrodynamic details he developed so lovingly were distractions—he failed to see the forest for the trees.

The general proof of $E = mc^2$ remained elusive until 1911, when Laue finally derived the mass-energy relation for an arbitrary closed "static" system, that is, a system with a time-independent energy-momentum tensor containing any electric, mechanical, elastic, chemical, thermal, etc. energies and stresses whatsoever (Laue, 1911). His proof exploited Minkowski's tensor formalism; it was concise and elegant, and avoided the tedious dynamical details that had frustrated Einstein. Laue simply took the energy-momentum tensor as his starting point and integrated its $T^{0\mu}$ components over the volume of the system to obtain the total energy and momentum; he also showed that for a closed static system the conservation law $\partial_k T^{k\mu} = 0$ implies that the volume integrals of the stresses $T^{kl}$ are zero in the rest frame. From the Lorentz-transformation properties of the components of the energy-momentum tensor he was then able to prove that the volume integrals of the $T^{0\mu}$ components of the energy-momentum tensor transform as a four vector, that is, the total energy and the total momentum of the system transform as a four-vector. This implies that the energy and the momentum are necessarily $E = E^0 / \sqrt{1 - v^2/c^2}$ and $\mathbf{p} = (E^0/c^2)\mathbf{v} / \sqrt{1 - v^2/c^2}$, where, in Laue's notation, $E^0$ is the energy in the center-of-energy frame (that is, the reference frame in which the momentum is zero). Thus, the energy, kinetic energy, and momentum of an extended system have exactly the same form as those of a particle, with a mass equal to the rest energy divided by $c^2$. In Laue's own words: "…a closed static system in uniform motion behaves like a point mass of rest-mass $m^0 = E^0/c^2$."

In 1918, as a by-product of an investigation of energy and momentum in general relativity, Klein achieved a generalization of Laue's proof (Klein, 1918a). He avoided Laue's restrictive assumption that in the rest frame of the system the energy-momentum tensor is time-independent; he merely assumed that the system is closed, with a conserved energy-momentum tensor. He integrated the general conservation law $\partial_\mu T^{\mu\alpha} = 0$ over the worldtube of the system and cleverly exploited Gauss' theorem in 4 D to show that Laue's result remains valid for a time-dependent energy momentum tensor.[7]

Einstein produced a variant of Laue's proof in an unpublished manuscript (Einstein, 1912) on the theory of relativity, written in 1912 and presumably sent to the publisher of the *Handbuch der Radiologie* at that time. There were long delays in the publication, and the *Handbuch* volume was finally printed in 1924 without Einstein's contribution, because he refused permission for his name to be attached to a revision of his manuscript prepared by his assistant Jacob Grommer. Einstein included this proof in his 1914-1915 Berlin lectures (Einstein, 1914a) and in his 1921 Princeton lectures, which were published in *The Meaning of Relativity* in the same year (Einstein, 1921); and he incorporated versions of this proof in two papers on general relativity in 1914 that tangentially touched on the mass-energy equivalence (Einstein, 1914b). In none of these writings did Einstein give any reference to Laue, but in view of the time line and the similarity of Einstein's mathematical arguments to those of Laue in their focus on the four-vector character of the energy-momentum integrals, there can be little doubt that Einstein's proofs were inspired by Laue's.

Einstein's variant of Laue's proof contains a flagrant and fatal mistake, recognized by Klein in a letter to Einstein in 1918 (Klein 1918b).[8] Instead of taking as starting point the conservation law $\partial_\mu T^{\mu\alpha} = 0$ in the absence of external forces, Einstein starts with the rate of change of the energy-momentum tensor in the presence of an external four-vector force density, $\partial_\mu T^{\mu\alpha} = f^\alpha$. He assumes that the force density $f^\alpha$ is nonzero over some limited time interval, and he shows, quite correctly, that the *change* in the energy-momentum of the system is then necessarily a four vector. He then asserts "Since the quantities themselves may be presumed to transform in the same way as their increments, we infer that the aggregate of the four quantities $I_x, I_y, I_z, iE$ [energy-momentum] has itself vector character…" (Einstein, 1921, p. 44). But this assertion is obviously false. Einstein is not entitled to *presume* what he wants to presume. That the "quantities themselves …transform in the same way as their increments" needs to be *proved*. All that Einstein proves by his argument is that the action of the force does not alter the four-vector character of the total energy-momentum of the system—he proves that the energy-momentum will be a four vector after the force acts if and only if it is a four vector before, and this tells us *nothing* whatsoever about whether the energy-momentum is a four vector or not.

This is an astonishing misstep, all the more so because Einstein was aware of it, but refused to recognize it for what it was. In a footnote in the 1912 manuscript, he highlighted this misstep: "To be sure, this is not rigorous, because additive constants might be present that do not have the character of a vector; but this seems so artificial that we will not dwell on this possibility at all." (Einstein, 1912, p. 158). Despite this admission, he persisted in this mistake in all the revisions and republications of this argument, over a span of more than forty years (he first wrote down this argument in the 1912 manuscript, included it in several papers on general relativity and in his 1921 book *The Meaning of Relativity*, and revised this book in four subsequent editions, with the last of these in 1955—and he never corrected his mistake.)[9]

In the two last papers on $E = mc^2$ in 1935 and 1946, Einstein reverted to the mistakes of his earliest papers. In the 1935 paper, which is the published version of the Josiah Willard Gibbs lecture he delivered in Philadelphia in 1934, he again assumed—without any justification—that the energy and momentum of a system have a particle-like dependence on velocity.[10]

And in the 1946 paper he repeated exactly the same mistake as in 1905, that is, he again dealt only with a low-speed approximation (but he now performed the calculation with the momentum change produced by the emission of two light pulses, rather than with the energy change, as in 1905).

Thus, the vulgar identification of Einstein's name with the equation $E = mc^2$ is not justified by the historical facts. Einstein does not have a solid claim on this equation, neither in terms of priority nor in terms of proof. Einstein himself thought otherwise. In 1907 he sent an irate letter to Stark complaining "I find it rather strange that you do not recognize my priority in the relationship of inertial mass and energy…"(Einstein, 1907c). Stark had credited Planck with this result, being apparently unaware of Einstein's 1905 paper. Stark replied in a conciliatory manner; had he known about Einstein's mistakes, he would have stood his ground.

Why did Einstein's name become so closely linked with $E = mc^2$? To a large extent this can be attributed to the influence he exerted by his repeated attempts to find a proof. These attempts, though defective, served to promote this equation and instigated other physicists to search for better proofs. However, as Mehra remarked in a letter to Wigner in a different context (Mehra, 1974, p. 86),[11] to some extent it must be attributed to "the sociology of science, the question of the cat and the cream. Einstein was the big cat of relativity, and the whole saucer of its cream belonged to him by right and by legend, or so most people assume!"

### Acknowledgments

I thank Dennis Clougherty and Peter Brown for helpful comments on an earlier version of this manuscript.

### References

Arzeliès, H. (1966), *Rayonnement et dynamique du corpuscule chargé fortement accéléré*. Paris: Gauthiers-Villars.

Einstein, A. (1905), Ist die Trägheit eines Körpers von seinem Energieinhalt abhängig? *Ann. d. Phys. 18*, 639-641. Translated in H. A. Lorentz et al. (1923), *The Principle of Relativity* (pp. 69-71). London: Methuen and Co.

Einstein, A. (1906), Das Prinzip von der Erhaltung der Schwerpunktsbewegung und die Trägheit der Energie, *Ann. d. Phys. 20*, 627-633.

Einstein, A. (1907a), Über die vom Relativitätsprinzip geforderte Trägheit der Energie, *Ann. d. Phys. 23*, 371-384.

Einstein, A. (1907b), Über das Relativitätsprinzip und die aus demselben gezogenen Folgungen, *Jahrbuch Radioaktivität 4*, 411-462.

Einstein, A. (1907c). In A. Einstein (1993), *The Collected Papers of Albert Einstein*, *Vol. 5* (Document 85). Princeton: Princeton University Press.

Einstein, A. (1912). In A. Einstein (2003), *Einstein's 1912 Manuscript on the Special Theory of Relativity*. New York: George Braziller Publishers. Also reprinted in A. Einstein (1995), *The Collected Papers of Albert Einstein*, *Vol. 4* (Document 1). Princeton: Princeton University Press.

Einstein, A. (1914a). In A. Einstein (1996), *The Collected Papers of Albert Einstein, Vol. 6* (Document 7). Princeton: Princeton University Press.

Einstein, A. (1914b). In A. Einstein (1995), *The Collected Papers of Albert Einstein, Vol. 4* (Documents 24 and 25). Princeton: Princeton University Press.

Einstein, A. (1921), *The Meaning of Relativity*. Princeton: Princeton University Press.

Einstein, A. (1935), Elementary Derivation of the Equivalence of Mass and Energy, *Bull. Am. Math. Soc. 41*, 223-230.

Einstein, A. (1946), An Elementary Derivation of the Equivalence of Mass and Energy, *Technion Yearbook 5*, 16-17. Reprinted in Einstein, A. (1967), *Out of My Later Years*. Totowa, NJ: Littlefield, Adams, & Co.

Einstein, A. (1987-2006), *The Collected Papers of Albert Einstein, Vols. 1-10*. Princeton: Princeton University Press.

Ives, H. E. (1952), Derivation of the Mass-Energy Relation, *Am. J. Phys. 42*, 540-543.

Jammer, M. (1961), *Concepts of Mass in Classical and Modern Physics*. Mineola, NY: Dover Publications.

Klein, F. (1918a), Über die Integralform der Erhaltungssätze und die Theorie der räumlich-geschlossenen Welt, *Nach. Gesells. Wissensch. Göttingen, Math.-Physik. Klasse*, 394-423.

Klein, F. (1918b). In A. Einstein (1998), *The Collected Papers of Albert Einstein, Vol. 8B* (Document 554). Princeton: Princeton University Press.

Laue, M. (1911), Zur Dynamik der Relativitätstheorie, *Ann. d. Phys. 35*, 524-542.

Miller, A. I. (1981), *Albert Einstein's Special Theory of Relativity*. Reading, MA: Addison-Wesley.

Medina, R. (2006), The inertia of stress, *Am. J. Phys. 74*, 1031-1034.

Mehra, J. (1974), *Einstein, Hilbert, and the Theory of Gravitation*. Dordrecht: Reidel.

Møller, C. (1952), *The Theory of Relativity*. Oxford: Clarendon Press.

Ohanian, H. C. and Ruffini, R. (1994), *Gravitation and Spacetime*. New York: W. W. Norton & Co.

Planck, M. (1908), Zur Dynamik bewegter Systeme, *Ann. d. Phys. 26*, 1-34.

Rohrlich, F. (1965), *Classical Charged Particles*. Reading, MA: Addison-Wesley.

Whittaker, E. (1960), *A History of the Theories of Aether and Electricity*. New York: Harper.

Stachel, J. and Torretti, R. (1982), Einstein's first derivation of the mass-energy equivalence, *Am J. Phys. 50*, 760-763.

---

[1] Such a non-universal behavior of the kinetic-energy functions of extended bodies was considered possible in 1905; and Ehrenfest even considered it possible that the kinetic energy might depend on the orientation of the body relative to its direction of motion. If it had been known that the kinetic energy functions are universal, then Einstein could have concluded that they must necessarily be of the form of Eq. (1), and then his argument would have been logically complete, although still affected by errors ii) and iii).

[2] Oddly, in the 1905 paper Einstein calls attention to the fact that, by Eq. (2), the *change* in the kinetic energy has the same dependence on $v$ as the kinetic energy of a particle, but he fails to see the implications of this comment.

[3] It should be kept in mind that, in 1905, not even the relativistic mechanics of particles was thought to be well established, neither theoretically or experimentally.

[4] If the changes generated by the forces in a light crossing time are small, then it is possible to define the kinetic energy approximately, that is, the extended system can then be treated approximately as a particle.

[5] Planck does not explain his remark, and we can therefore not be sure exactly what motivated it. Maybe he was concerned with the problem of time-dependent forces, but Stachel-Coretti contend that Planck's remark was motivated by the inconsistencies that he had identified in the kinetic energy of a moving volume of blackbody radiation. However, it is likely that Planck was aware that these inconsistencies are also present in other systems, such as a volume of ideal, relativistic gas. It is also possible that Planck meant to say that the simple construction of the translational kinetic that can be performed in nonrelativistic physics (by introducing the velocity of the center mass) cannot be transcribed into relativistic physics.

[6] The centroid of the energy distribution is frame-dependent, that is, it is not a four vector. Centroids calculated in different inertial reference frames differ by a time-independent displacement proportional to the spin angular momentum of the body. It can be shown that for an isolated system, the translational velocity of all these centroids of energy coincides with the velocity of the "center-of-momentum" frame, that is, the reference frame in which the momentum of the system is zero; see Møller (1952), Section 64.

[7] Full details are given in Møller, 1952, Section 63. For a concise treatment, see Ohanian and Ruffini, 1994, pp. 87-90.

[8] This letter led to an exchange of further letters between Klein and Einstein with various erroneous attempts at proofs, until Klein finally produced the correct proof given in Klein, 1918a.

---

[9] The 1912 manuscript also presents another, alternative proof of the mass-energy relation [Einstein, 1912, p. 108], which relies on an examination of the process of emission of two light pulses by a system, as in the 1905 paper. In contrast to 1905, Einstein now assumes explicitly that the system can be regarded as particle-like and that its energy is $E = Mc^2 / \sqrt{1 - v^2 / c^2}$ . This is a genuine error of circular reasoning, because if this formula for the energy is assumed known, then there is nothing left to prove—for $v = 0$, the formula yields the energy $E = Mc^2$, QED. However, since Einstein never *published* this "proof," it would be unfair to hold it against him.

[10] He tries to camouflage this mistake by a semantic quibble, calling the systems under consideration "particles." However, these so-called particles are assumed to be capable of absorbing and storing energy, which means they cannot be structureless, pointlike entities—they are necessarily physical *systems* with internal structures of some finite extent. At best, we might assume that theses systems are very small, and therefore approximately pointlike. However, no matter how small, a system will reveal its non-pointlike features if the force field acting on it varies in space on a scale smaller than the size of the system or in time on a scale shorter than the light-crossing time.

[11] Mehra was addressing the Einstein-Hilbert controversy over the priority of discovery of the "Einstein" equations for the gravitational field. But Einstein has a much better claim on these equations than on the mass-energy relation.